\documentclass[sn-mathphys,Numbered]{sn-jnl}

\usepackage[table]{xcolor}
\usepackage{float}

\usepackage{graphicx}%
\usepackage{multirow}%
\usepackage{amsmath,amssymb,amsfonts}%
\usepackage{amsthm}%
\usepackage{mathrsfs}%
\usepackage{cleveref}
\usepackage[title]{appendix}%
\usepackage{xcolor}%
\usepackage{textcomp}%
\usepackage{manyfoot}%
\usepackage{booktabs}%
\usepackage{algorithm}%
\usepackage{algorithmicx}%
\usepackage{algpseudocode}%
\usepackage{listings}%

\usepackage[nolist,nohyperlinks]{acronym}
\begin{acronym}
    \acro{RS}{Recommender System}
    \acro{CF}{Collaborative Filtering}
    \acro{MF}{Matrix Factorization}
    \acro{MAE}{Mean Absolute Error}
    \acro{BiasedMF}{Biased Matrix Factorization}
    \acro{NDCG}{Normalized Discounted Cumulative Gain}
    \acro{BeMF}{Bernoulli Matrix Factorization}
    \acro{KNN}{K Nearest Neighbors}
    \acro{PMF}{Probabilistic Matrix Factorization}
    \acro{NMF}{non-Negative Matrix Factorization}
    \acro{BNMF}{Bayesian non-Negative Matrix Factorization}
    \acro{URP}{User Ratings Profile Model}
    \acro{NN}{Neural Network}
    \acro{NCF}{Neural Collaborative Filtering}
\end{acronym}

\begin{document}

\title{Comprehensive Evaluation of Matrix Factorization Models for Collaborative Filtering Recommender Systems}

\author{\fnm{Jesús} \sur{Bobadilla}}\email{jesus.bobadilla@upm.es}
\author{\fnm{Jorge} \sur{Dueñas-Lerín}}\email{jorgedl@alumnos.upm.es}
\author{\fnm{Fernando} \sur{Ortega}}\email{fernando.ortega@upm.es}
\author{\fnm{Abraham} \sur{Gutiérrez}}\email{abraham.gutierrez@upm.es}

\affil{\orgdiv{ETSISI}, \orgname{Technical University of Madrid}, \orgaddress{\street{Ctra. de Valencia Km. 7}, \city{Madrid}, \country{Spain}}}


\abstract{Matrix factorization models are the core of current commercial collaborative filtering Recommender Systems. This paper tested six representative matrix factorization models, using four collaborative filtering datasets. Experiments have tested a variety of accuracy and beyond accuracy quality measures, including prediction, recommendation of ordered and unordered lists, novelty, and diversity. Results show each convenient matrix factorization model attending to their simplicity, the required prediction quality, the necessary recommendation quality, the desired recommendation novelty and diversity, the need to explain recommendations, the adequacy of assigning semantic interpretations to hidden factors, the advisability of recommending to groups of users, and the need to obtain reliability values. To ensure the reproducibility of the experiments, an open framework has been used, and the implementation code is provided.}

\maketitle

\section{Introduction}
\ac{RS} \cite{batmaz2019review} is the field of artificial intelligence specialized in user personalization. Mainly, \acp{RS} provide accurate item recommendations to users: movies, trips, books, music, etc. Recommendations are made following some filtering approach. The most accurate filtering approach is the \ac{CF} \cite{bobadilla2020deep}, where recommending to an active user involves a first stage to make predictions about all his or her not consumed or voted items. Then, the top predicted items are recommended to the active user. The \ac{CF} approach assumes the existence of a dataset that contains explicitly voted items or implicitly consumed items from a large number of users. Remarkable commercial \acp{RS} are Amazon, Spotify, Netflix, or TripAdvisor.

Regardless of the machine learning model used to implement \ac{CF}, the key concept is to extract user and item patterns and then to recommend to the active user those items that he or she has not voted or consumed, and that similar users have highly valued. It fits with the \ac{KNN} memory-based algorithm \cite{zhu2018efficient}, and it is the reason why the initial RS research was based on \ac{KNN}. There are also some other filtering approaches such as demographic, social, content-based, context-aware, and their ensembles. Demographic filtering \cite{bobadilla2021deepfair} makes use of user information such as gender, age, or zip code, and item information such as movie genre, country to travel, etc. Social filtering \cite{carbo2005,medel2022} has a growing importance in current RS, due to the social networks boom. The existence of trust relations and graphs \cite{caro2021} can improve the quality of the CF recommendations. In this decentralized and dynamic environment, trust between users provides additional information to the centralized set of ratings. Trust relationships can be local, collective, or global \cite{afef2016}; local information is based on shared users’ opinions, collective information uses friends’ opinions, whereas global information relates to users’ reputation \cite{pinyol2013}. Content-based filtering \cite{deldjoo2020recommender} recommends items with the same type (content) to consumed items (e.g. to recommend Java books to a programmer that bought some other Java book). Context-aware filtering \cite{kulkarni2020context} uses GPS information, biometric sensor data, etc. Finally, ensemble architectures \cite{forouzandeh2021presentation} get high accuracy by merging several types of filtering.

Memory-based algorithms have two main drawbacks: their accuracy is not high, and each recommendation process requires to recompute the whole dataset. Model-based approaches solve both problems: their accuracy is higher than that of memory-based methods, and they first create a model from the dataset. From the created model we can make many different recommendations, and it can be efficiently updated when the dataset changes. \ac{MF} \cite{mnih2007probabilistic} is the most popular approach to implement current \acp{RS}: it provides accurate predictions, it is conceptually simple, it has a straightforward implementation, the model learns fast, and also updates efficiently. The \ac{MF} model makes a compression of information, coding very sparse and large vectors of discrete values (ratings) to low dimensional embeddings of real numbers, called hidden factors. The hidden factors, both from the user vector and from the item vector, are combined by means of a dot product to return predictions. This is an iterative process in which the distance between training predictions and their target ratings is minimized. 

The \ac{PMF} model based on \ac{MF} \cite{mnih2007probabilistic} scales linearly with the size of the data set. It also returns accurate results when applied to sparse, large, and imbalanced \ac{CF} datasets. \ac{PMF} has also been extended to include an adaptive prior on the model parameters, and it can generalize adequately, providing accurate recommendation to cold-start users. \ac{CF} \acp{RS} are usually biased. A typical \ac{CF} bias source comes from the fact that some users tend to highly rate items (mainly 4 and 5 stars), whereas some other users tend to be more restrictive in their ratings (mainly 3 and 4 stars). This fact leads to the extension of the \ac{MF} model to handle biased data. An user-based rating centrality and an item-based rating centrality \cite{wu2018optimization} have been used to improve the accuracy of the regular \ac{PMF}. These centrality measures are obtained by processing the degree of deviation of each rating in the overall rating distribution of the user and the item. \ac{NMF} \cite{fevotte2011algorithms} can extract significant features from sparse and non-negative \ac{CF} datasets (please note that \ac{CF} ratings are usually a non-negative number of stars, listened songs, watched movies, etc.). When nonnegativity is imposed, prediction errors are reduced and the semantic interpretability of hidden factors is easier. The \ac{BeMF} \cite{ortega2021providing} has been designed to provide both prediction and reliability values; this model uses the Bernoulli distribution to implement a set of binary classification approaches. The results of the binary classification are combined by means of an aggregation process. The \ac{BNMF} \cite{hernando2016non} was designed to provide useful information about user groups, in addition to the \ac{PMF} prediction results. The authors factorize the rating matrix into two nonnegative matrices whose components lie within the range [0, 1]. The resulting hidden factors provide an understandable probabilistic meaning. Finally, The \ac{URP} is a generative latent variable model \cite{marlin2003modeling}; it produces complete rating user profiles. In the \ac{URP} model, first attitudes for each item are generated, then a user attitude for the item is selected from the set of existing attitudes. \ac{URP} borrows several concepts from LDA \cite{blei2003latent} and the multinomial aspect model \cite{hofmann2001learning}.

The set of \ac{MF} models mentioned above: \ac{PMF}, \ac{BiasedMF}, \ac{NMF}, \ac{BeMF}, \ac{BNMF}, and \ac{URP}, can be considered representative in the \ac{CF} area. These models will be used in this paper to compare their behavior when applied to representative datasets. Specifically, the following quality measures will be tested: \ac{MAE}, novelty, diversity, precision, recall, and \ac{NDCG}. Prediction accuracy will be tested using \ac{MAE} \cite{gunawardana2015evaluating}, whereas \ac{NDCG}, Precision and Recall \cite{aggarwal2016evaluating} will be used to test recommendation accuracy. Modern \ac{CF} models should be tested not only regarding accuracy, but also beyond accuracy properties \cite{bobadilla2022}: novelty \cite{vargas2011rank,castells2011novelty} and diversity \cite{vargas2011intent}. Novelty can be defined as the quality of a system to avoid redundancy; diversity is a quality that helps to cope with ambiguity or under-specification. The models have been tested using four \ac{CF} datasets: \texttt{MovieLens} (100K and 1M versions) \cite{harper2015movielens}, \texttt{Filmtrust} \cite{golbeck2006filmtrust} and \texttt{MyAnimeList} \cite{miller2014recommender}. These are representative open datasets and are popular in RS research. 

Overall, this paper provides a complete evaluation of \ac{MF} methods, where the \ac{PMF}, \ac{BiasedMF}, \ac{NMF}, \ac{BeMF}, \ac{BNMF}, and \ac{URP} models have been tested using representative \ac{CF} quality measures, both for prediction and recommendation, and also beyond accuracy ones. As far as we know this is the experimental most complete work evaluating current \ac{MF} models in the \ac{CF} area.

The rest of the paper is structured as follows: \Cref{methods} introduces the tested models, the experiment design, the selected quality measures, and the chosen datasets. \Cref{results} shows the obtained results and provides their explanations \Cref{discussion}. \Cref{conclusions} highlights the main conclusions of the paper and the suggested future works. Finally, a references section lists current research in the area.

\section{Methods and Experiments}\label{methods}

This section abstracts the fundamentals of each baseline model (\ac{PMF}, \ac{BiasedMF}, \ac{NMF}, \ac{BeMF}, \ac{BNMF}, \ac{URP}), introduces the tested quality measures (\ac{MAE}, precision, recall, \ac{NDCG}, novelty, diversity), and shows the main parameters of the tested datasets (\texttt{Movielens}, \texttt{FilmTrust}, \texttt{MyAnimeList}). Experiments are performed by combining the previous entities.

The vanilla \ac{MF} \cite{mnih2007probabilistic,koren2009matrix} is used to generate rating predictions from a matrix of ratings $R$. This matrix contains the set of casted ratings (explicit or implicit) from a set of users $U$ to a set of items $I$. Since regular users only vote or consume a very limited subset of the available items, matrix $R$ is very sparse. The \ac{MF} key concept is to compress the very sparse item and user vectors of ratings to small size and dense item and user vectors of real numbers; these small size dense vectors can be considered as embeddings, and they usually are called ‘hidden factors’, since each embedding factor codes some complex non-lineal (‘hidden’) relation of user or item features. The parameter $K$ is usually chosen to set the embedding (hidden factors) size. \ac{MF} makes use of two matrices: $P(|{U}| * K)$ to contain the $K$ hidden factors of each user, and $Q(|I| * K)$ to contain the $K$ hidden factors of each item. To predict how much a user $u$ likes an item $i$, we compare each hidden factor of $u$ with each corresponding hidden factor of $i$. Then, the dot product $u \cdot i$ can be used as suitable \ac{CF} prediction measure. \ac{MF} predicts ratings by minimizing errors between the original $R$ matrix and the predicted $\hat{R}$ matrix:
\begin{align}
    R &\approx P \times Q^{T} = \hat{R}\\
    \hat{r}_{ui} &= p_{u} \cdot q_{i}^{T} = \sum^{K}_{k=1}p_{uk} \cdot q_{ki}
\end{align}

Using gradient descent, we minimize learning errors (differences between real ratings $r$ and predicted ratings $\hat{r}$).
\begin{equation}
    \label{eq:error_minimization}
    e^{2}_{ui} = (r_{ui} - \hat{r}_{ui})^{2} = (r_{ui} - \sum^{K}_{k=1}p_{uk} \cdot q_{ki} )^{2}
\end{equation}

To minimize the error, we differentiate equation \Cref{eq:error_minimization} with respect to $p_{uk}$ and $q_{ki}$:
\begin{align}
    \frac{\partial}{\partial p_{uk}}e^{2}_{ui} &= - 2(r_{ui} - \hat{r}_{ui})q_{ki} = - 2e_{ui}q_{ki}\\
    \frac{\partial}{\partial q_{ki}}e^{2}_{ui} &= - 2(r_{ui} - \hat{r}_{ui})p_{uk} = - 2e_{ui}p_{uk}
\end{align}

Introducing the learning rate $\alpha$, we can iteratively update the required hidden factors $p_{uk}$ and $q_{ki}$:
\begin{align}
    p'_{uk} &= p_{uk} + \alpha \frac{\partial}{\partial p_{uk}}e^{2}_{ui} = p_{uk} + 2 \alpha e_{ui}q_{ki}\\
    q'_{ki} &= q_{ki} + \alpha \frac{\partial}{\partial q_{ki}}e^{2}_{ui} = q_{ki} + 2 \alpha e_{ui}p_{uk}
\end{align}

\ac{CF} datasets have biases, since different users vote or consume items in different ways. In particular, there are users who are more demanding than others when rating products or services. Analogously, there are items more valued than others on average. Biased MF \cite{wu2018optimization} is designed to consider data biases; The following equations extend the previous ones, introducing the bias concept and making the necessary regularization to maintain hidden factor values in their suitable range:
\begin{equation}
    \hat{r}_{ui} = \mu + b_{u} + b_{i} + p_{u} \cdot q_{i}^{T}
\end{equation}
where $\mu, b_{u}, b_{i}$ are the average bias, the user bias and the item bias.

We minimize the regularized squared error:
\begin{equation}
    \sum_{r_{ui} \in R_{train}}(r_{ui} - \hat{r}_{ui})^{2} + \lambda (b_{i}^{2} + b_{u}^{2} + ||{q_{i}}||^{2} + ||{p_{u}}||^{2})
\end{equation}
where $\lambda$ is the regularization term.

Obtaining the following updating rules:
\begin{align}
    b'_{u} &= b_{u} + \alpha (e_{ui} - \lambda b_{u})\\
    b'_{i} &= b_{i} + \alpha (e_{ui} - \lambda b_{i})\\
    p'_{u} &= p_{u} + \alpha (e_{ui} \cdot q_{i} - \lambda p_{u})\\
    q'_{i} &= q_{i} + \alpha (e_{ui} \cdot p_{u} - \lambda q_{i})
\end{align}

\ac{NMF} \cite{fevotte2011algorithms} can be considered as a regular \ac{MF} subject to the following constraints:
\begin{equation}
    R \geq 0, P \geq 0, Q \geq 0
\end{equation}

In the \ac{NMF} case, predictions are made by linearly combining positive coefficients (hidden factors). \ac{NMF} hidden factors are easier to semantically interpret than regular \ac{MF} ones: sometimes it is not straightforward to assign semantic meanings to negative coefficient values. In the \ac{CF} context, another benefit of using \ac{NMF} decomposition is the emergence of a natural clustering of users and items. Intuitively, users and items can be clustered according to the dominant factor (i.e. the factor having the highest value). In the same way, the original features (gender, age, item type, item year, etc.) can be grouped according to the factor (from the k hidden factors) on which they have the greatest influence. This is possible due to the condition of positivity of the coefficients.

\ac{BeMF} \cite{ortega2021providing} is an aggregation-based architecture that combines a set of Bernoulli factorization results to provide pairs $<$prediction, reliability$>$. \ac{BeMF} uses as many Bernoulli factorization processes as possible scores in the dataset. Reliability values can be used to detect shilling attacks, to explain the recommendations, and to improve prediction and recommendation accuracy \cite{bobadilla2021neural}. \ac{BeMF} is a classification model based on the Bernoulli distribution. It adequately adapts to the expected binary results of each of the possible scores in the dataset. Using \ac{BeMF}, the prediction for user $u$ to item $i$ is a vector of probabilities $(p^{1}_{ui}, ..., p^{D}_{ui})$, where $p^{s}_{ui}$ is the probability that $i$ is assigned the s-th score from user $u$. The \ac{BeMF} model can be abstracted as follows:

Let $S = \{s_{1}, ..., s_{D}\}$ be the set of D possible scores in the dataset (e.g. 1 to 5 stars: D=5). From R we generate D distinct matrices $(R^{s1}, ..., R^{sD})$; each $R^{s} = (R^{s}_{ui})$ matrix is a sparse matrix such that $R^{s}_{ui} = 1$. BeMF will attempt to fit the matrices $R^{s1}, ..., R^{sD}$ by performing $D$ parallel \acp{MF}.

The \ac{BeMF} assumes that, given the user $P$ matrix and the item $Q$ matrix containing $k>0$ hidden factors, the rate $R_{ui}$ is a Bernoulli distribution with the success probability $\psi (P_{u} \cdot Q_{i})$. The mass function of this random variable is:
\begin{equation}
    p(R_{ui}|P_{u}, Q_{i}) = \begin{cases} 
      \psi (P_{u}Q_{i}) & if R_{ui} = 1,\\
      1- \psi (P_{u}Q_{i}) & if R_{ui} = 0 \\
   \end{cases}
\end{equation}

The associated likelihood is:
\begin{equation}
    \ell(R|UV) = \sum_{R_{ui}=1} log(\psi (P_{u}Q_{i})) +   \sum_{R_{ui}=0} log(1 - \psi (P_{u}Q_{i})) 
\end{equation}

The \ac{BeMF} updating equations are:
\begin{equation}
    \begin{split}
        P'_{u} = & P_{u} + \gamma(\sum_{\{i|R_{ui}=1\}}(1 - logit(P_{u}Q_{i}))Q_{i}\\ 
        & + \sum_{\{i|R_{ui}=1\}} logit(P_{u}Q_{i})Q_{i} - \eta P_{u})
    \end{split}
\end{equation}
\begin{equation}
    \begin{split}
        Q'_{i} = & Q_{i} + \gamma(\sum_{\{i|R_{ui}=1\}}(1 - logit(P_{u}Q_{i}))P_{u}\\
        & + \sum_{\{i|R_{ui}=1\}} logit(P_{u}Q_{i})P_{u} - \eta Q_{i})
    \end{split}
\end{equation}

And the aggregation to obtain the final output $\Phi$:
\begin{equation}
    \Phi (u,i) = \frac{1}{\sum^{s}_{\alpha = 1} \psi (P^{S_{\alpha}}_{u}Q^{S_{\alpha}}_{i})}(\psi (P^{S_{1}}_{u}Q^{S_{1}}_{i}), ..., \psi (P^{S_{D}}_{u}Q^{S_{D}}_{i}))
\end{equation}
where $\Phi (u,i) = (p^{1}_{ui}, ..., p^{D}_{ui}), 0 \leq p^{\alpha}_{ui} \leq 1, \sum_{\alpha}p^{\alpha}_{ui} = 1$. Let $\alpha_{0} = argmax_{\alpha}p^{\alpha}_{u}$; the prediction is: $\hat{R}_{ui} = s_{\alpha_{0}}$, and the reliability is $p^{\alpha_{0}}_{ui}$.

\ac{BNMF} \cite{hernando2016non} provides a Bayesian-based \ac{NMF} model that not only allows accurate prediction of user ratings, but also to find groups of users with the same tastes, as well as to explain recommendations. The \ac{BNMF} model approximates the real posterior distribution $\rho (\emptyset_{u}, k_{ik}, z_{ui} | \rho_{ui})$ by the distribution:
\begin{equation}
    \begin{split}
    q (\emptyset_{u}, k_{ik}, z_{ui}) = & \Pi^{N}_{u=1}q_{\emptyset_{u}}(\emptyset_{u}) \Pi^{M}_{i=1} \Pi^{K}_{k=1}q_{k_{ik}}(k_{ik})\\
    & \Pi_{r_{ui}\neq \ast}q_{z_{ui}}(z_{ui})
    \end{split}
\end{equation}
where:
\begin{itemize}
    \item $z_{ui} \sim Cat(\emptyset_{u})$ is a random variable from a categorical distribution.
    \item $\rho_{ui} \sim Bin(R, k_{i, z_{ui}})$ is a random variable from a Binomial distribution (which takes values from 0 to D-1).
    \item $p_{ui} = \sum_{k=1 ... K}a_{uk} \cdot b_{ik}$ (a and b are hidden matrices).
    \item $q_{ui} = \begin{cases}
            1 & if \ \  0 \leq p_{ui} < 0.2 \\
            2 & if \ \  0.2 \leq p_{ui} < 0.4 \\
            etc. & \\
         \end{cases}$
    \item $q_{\emptyset_{u}}(\emptyset_{u}) \sim Dir(\gamma_{u1}, ..., \gamma_{uk})$ follows a Dirichlet distribution.
    \item $q_{k_{ik}}(k_{ik}) \sim Beta(\epsilon^{+}_{ik}, \epsilon^{-}_{ik},)$ follows a Beta distribution.
    \item $q_{z_{ui}}(z_{ui}) \sim Cat(\lambda_{ui1}, ..., \lambda_{uik})$ follows a categorical distribution.
    \item $\lambda_{uik}$ are parameters to be learned: $\lambda_{ui1} + ... + \lambda_{uik} = 1$ 
\end{itemize}

\ac{BNMF} iteratively approximates parameters $\{\gamma_{uk}, \epsilon^{+}_{ik}, \epsilon^{-}_{ik}, \lambda_{uik} \}$:
\begin{align}
    \gamma_{uk} = & \alpha + \sum_{\{i|r_{ui}\neq\ast\}}\lambda_{uik}\\
    \epsilon^{+}_{ik} = & \beta + \sum_{\{i|r_{ui}\neq\ast\}}\lambda_{uik} \cdot r^{+}_{ui}\\
    \epsilon^{-}_{ik} = & \beta + \sum_{\{i|r_{ui}\neq\ast\}}\lambda_{uik} \cdot r^{-}_{ui}\\
    \begin{split}
        \lambda'_{uik} = & exp(\psi(\gamma_{uk}) + r^{+}_{ui}\cdot\psi(\epsilon^{+}_{ik}) + r^{-}_{ui}\cdot\psi(\epsilon^{-}_{ik})\\
        & - (D - 1) \cdot\psi(\epsilon^{+}_{ik} + \epsilon^{-}_{ik}))
    \end{split}\\
    r^{+}_{ui} = & \rho_{ui} = (D - 1)\cdot r^{\ast}_{ui}\\
    r^{-}_{ui} = & (D - 1) - \rho_{ui} = (D - 1)\cdot (1 - r^{\ast}_{ui})\\
    r^{\ast}_{ui} = & \frac{\rho_{ui}}{(D - 1)}
\end{align}
where $\psi$ is the digamma function as the logarithmic derivative of the gamma function.

\ac{URP} is a generative latent variable model \cite{marlin2003modeling}. The model assigns to each user a mixture of user attitudes. Mixing is performed by a Dirichlet random variable:

\begin{align}
    \begin{split}
        &P(\theta, z|\alpha, \beta, r^{u}) \approx q(\theta, z|\gamma^{u}, \emptyset_{u})\\
        & = q(\theta| \gamma^{u})\Pi^{M}_{y=1}q(Z_{y}=z_{y}| \emptyset^{u}_{y})
    \end{split}\\
    &\emptyset^{u}_{zy} \approx \Pi^{V}_{v=1}\beta^{\delta (r^{u}_{y},v)}_{vyz} exp (\psi (\gamma^{u}_{z}) - \psi (\sum^{k}_{j=1}\gamma^{u}_{j}))\\
    &\gamma^{u}_{z} = \alpha_{z} + \sum^{M}_{y=1} \emptyset^{u}_{zy}\\
    &\beta_{vyz} \approx \sum^{N}_{u=1} \emptyset^{u}_{zy} \delta (r^{u}_{y},v)\\
    &\psi (\alpha_{z}) = \psi(\sum^{K}_{z=1}\alpha_{z}) + \frac{1}{N} \cdot (\sum^{N}_{u=1}\psi(\gamma^{u}_{z})-\psi(\sum^{N}_{u=1}\gamma^{u}_{z}))\\
    &\alpha_{z} = \psi^{-1}(\psi (\alpha_{z}))
\end{align}

In this paper, baseline models will be tested using a) prediction measure, b) recommendation measures, and c) beyond accuracy measures. The chosen prediction measure is the \ac{MAE}, where the absolute differences of the errors are averaged. Absolute precision and relative recall measures are tested to compare the quality of an unordered list of N recommendations. The ordered lists of recommendations will be compared using the \ac{NDCG} quality measure. From the beyond accuracy metrics, we have selected novelty and diversity. Novelty returns the distance from the items the user ‘knows’ (has voted or consumed) to his recommended set of items. Diversity tells us about the distance between the set of recommended items. Recommendations with high novelty values are valuable, since they show to the user unknown types of items. Diverse recommendations are valuable because they provide different types of items (and each type of item can be novel, or not, to the user).

The GroupLens research group \cite{harper2015movielens} made available several CF datasets, collected over different intervals of time. MovieLens 100K and MovieLens 1M describe 5-star rating and free-text tagging activity. These data were created from 1996 to 2018. In the Movielens 100K dataset, users were selected at random from those who had rated at least 20 movies, whereas the MovieLens 1M dataset has not this constraint. Only movies with at least one rating or tag are included in the dataset. No demographic information is included. Each user is represented by an 'id', and no other information is provided. The dataset files are written as comma-separated values files with a single header row. Columns that contain commas (,) are escaped using double-quotes ("). These files are encoded as UTF-8. All ratings are contained in the file named 'ratings.csv'. Each line of this file after the header row represents one rating of one movie by one user, and has the following format: 'userId, movieId, rating, timestamp'. The lines within this file are ordered first by 'userId', then, within user, by 'movieId'. Timestamps represent seconds since midnight Coordinated Universal Time (UTC) of January 1, 1970. FilmTrust is a small dataset crawled from the entire FilmTrust website in June, 2011. As the Movielens datasets, it contains ratings voted from users to items; additionally, it provides social information structured as a graph network. Finally, MyAnimeList contains information about anime and 'otaku' consumers (anime, manga, video games and computers). Each user is able to add 'animes' to their completed list and give them a rating; this data set is a compilation of those ratings. The MyAnimeList CF information is contained in the file 'Anime.csv', where their main columns are 'anime\_id': myanimelist.net's unique 'id' identifying an anime; 'name':  full name of anime; 'genre': comma separated list of genres for this anime; 'type': movie, TV, OVA, etc; 'episodes': how many episodes in this show; 'rating': average rating out of 10 for this anime. These datasets are available in the Kaggle and GitHub repositories, as well as in the KNODIS research group CF4J \cite{ortega2018cf4j} repository  \href{https://github.com/ferortega/cf4j}{https://github.com/ferortega/cf4j}.

\Cref{tab:main_parameters} contains the values of the main parameters of the selected \ac{CF} data sets: \texttt{Movielens 100K}, \texttt{Movielens 1M}, \texttt{FilmTrust} and \texttt{MyAnimeList}. We have run the explained \ac{MF} models on each of the four \Cref{tab:main_parameters} datasets, testing the chosen quality measures. Please note that the \texttt{MyAnimeList} dataset ratings range from 1 to 10, whereas \texttt{MovieLens} datasets range from 1 to 5 and \texttt{FilmTrust} ranges from 0 to 5 with 0.5 increments. It is also remarkable the sparsity difference between \texttt{FilmTrust} and the rest of the tested datasets.

\begin{table}[ht]
    \caption{Main parameter values of the tested datasets}
    \label{tab:main_parameters}%
    \begin{tabular}{@{}llllll@{}}
        \toprule
        Dataset & \#users & \#items & \#ratings & Scores & Sparsity \\
        \midrule
        MovieLens100k & 943 & 1682 & 99,831 & 1 to 5 & 93.71 \\
        MovieLens1M & 6,040 & 3,706 & 911,031 & 1 to 5 & 95.94 \\
        MyAnimeList & 19,179 & 2,692 & 548,967 & 1 to 10 & 98.94 \\
        FilmTrust & 1,508 & 2,071 & 35,497 & 0 to 5 & 87.98 \\
        \botrule
    \end{tabular}%
\end{table}

Experiments have been performed using random search and applying four-fold cross-validation. To ensure reproducibility, we used a seed in the random process. Results shown in the paper are the average of the partial results obtained by setting the number $k$ of latent factors to $\{4, 8, 12\}$, and the number of \ac{MF} iterations to $\{25, 50, 75, 100\}$. Additionally, to run the \ac{PMF}, BiasedMF, and \ac{BeMF} models, both the learning rate and the regularization parameters have been set to $\{0.001, 0.01, 0.1, 1.0\}$. The \ac{BNMF} model requires two specific parameters: $\alpha$ and $\beta$ ; the chosen values por these parameters are: $\alpha$ = $\{0.2, 0.4, 0.6, 0.8\}$, and $\beta$ = $\{5, 15, 25\}$. The tested number of recommendations N ranges from 1 to 10. We have used 4 stars as recommendation threshold $\theta$ for datasets whose ratings range from 1 to 5, while the testing threshold has been 8 when \texttt{MyAnimeList} was chosen. The experiments have been implemented using the open framework \cite{ortega2021cf4j} and the code has been made available at  \url{https://github.com/KNODIS-Research-Group/choice-of-mf-models}

\section{Results}\label{results}

The prediction quality obtained by testing each baseline model is shown in \Cref{tab:prediction_quality}. The bold numbers correspond to the best results, and, of them, those highlighted gray are the top ones. As can be seen, BiasedMF and \ac{BNMF} models provide the best \ac{CF} prediction results. \ac{PMF}, \ac{NMF}, \ac{BeMF} and \ac{URP} seem to be more sensitive to the type of \ac{CF} input data.

\begin{table}[h]
    \caption{Prediction quality results using the Mean Absolute Error (MAE). The lower the error value, the better the result.}
    \label{tab:prediction_quality}%
    \begin{tabular}{@{}lllllll@{}}
        \toprule
         & PMF & BiasedMF & NMF & BeMF & BNMF & URP \\
        \midrule
        MovieLens 100K & 0.770 & \textbf{0.754} & 0.804 & 0.805 & \cellcolor{black!25}\textbf{0.748} & 0.837 \\
        MovieLens 1M & 0.729 & \textbf{0.712} & 0.744 & 0.748 & \cellcolor{black!25}\textbf{0.693} & 0.795 \\
        FilmTrust & 0.863 & \cellcolor{black!25}\textbf{0.652} & 0.876 & 0.712 & \textbf{0.666} & 0.831 \\
        MyAnimeList & 1.110 & \cellcolor{black!25}\textbf{0.926} & 1.147 & 1.034 & \textbf{0.943} & 1.159 \\
        \botrule
    \end{tabular}
\end{table}

\begin{figure}[ht]
    \centering
    \includegraphics[width=1.0\textwidth]{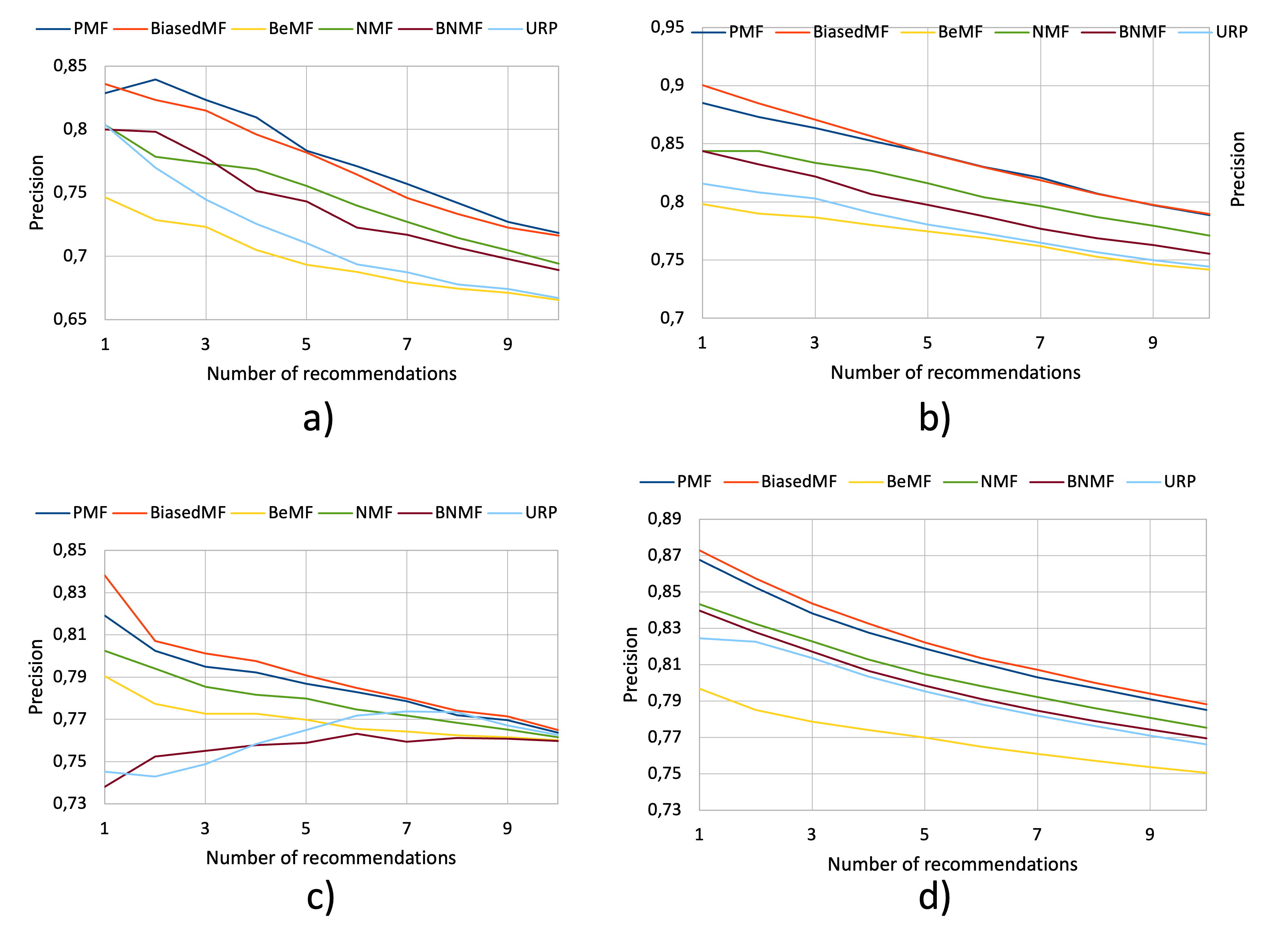}
    \caption{Precision recommendation quality results; a) \texttt{MovieLens100K}, b) \texttt{MovieLens 1M}, c) \texttt{FilmTrust}, d) \texttt{MyAnimeList}. The higher the values, the better the results.}
    \label{fig:precision_results}
\end{figure}

\Cref{fig:precision_results} shows the quality of recommendation obtained using the Precision measure. The most remarkable in Fig. \Cref{fig:precision_results} is the superiority of the models \ac{PMF} and BiasedMF. For the remaining models, \ac{URP} and \ac{BeMF} provide the worst results, whereas the nonnegative \ac{NMF} and \ac{BNMF} return an intermediate quality.  It is important to highlight the good performance of the BiasedMF model for both the prediction and the recommendation tasks.

\begin{figure}[H]
    \centering
    \includegraphics[width=0.8\textwidth]{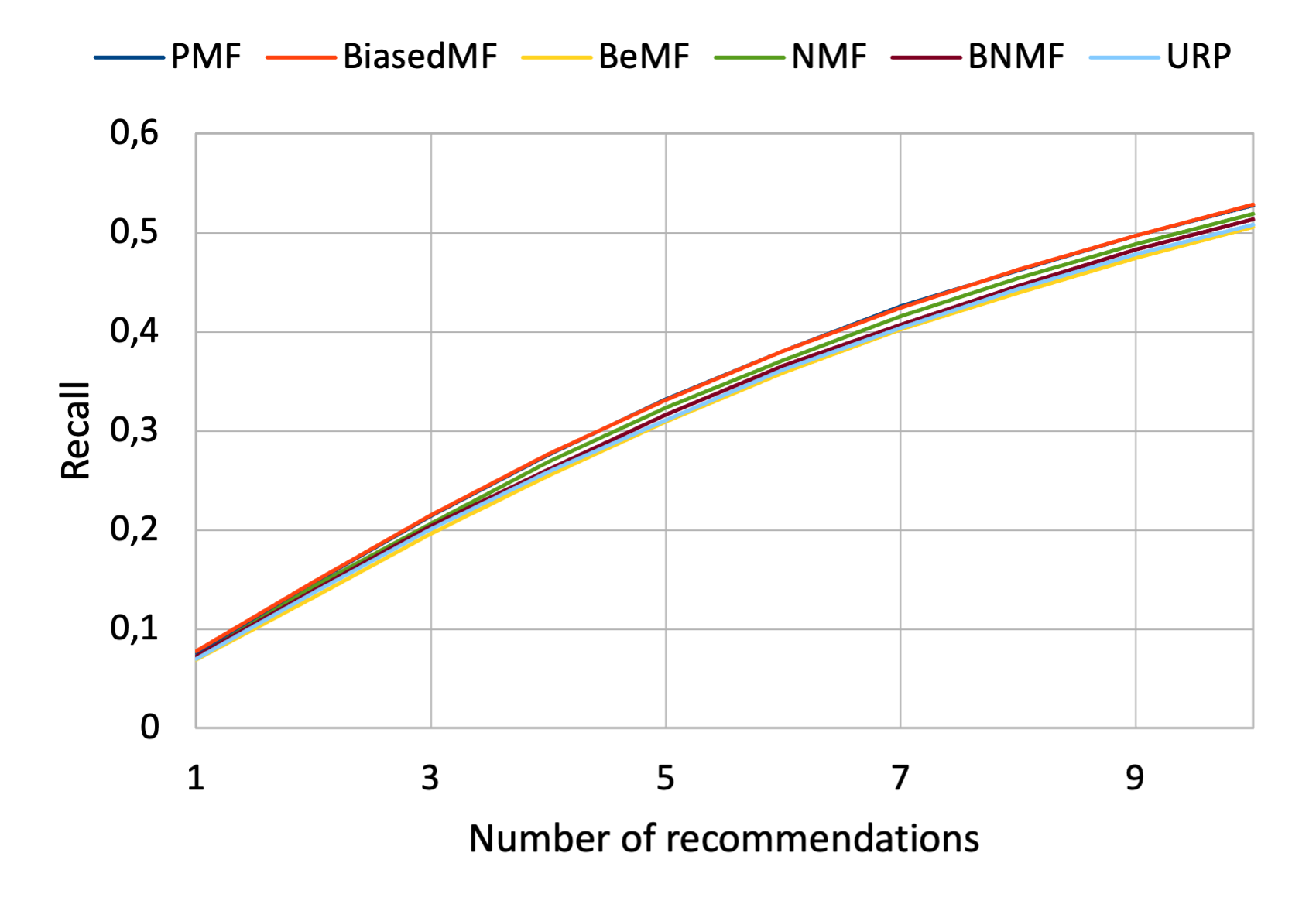}
    \caption{Recall Recommendation quality results obtained in the \texttt{MovieLens 1M} dataset. The results of the other three considered datasets are very similar to this one; to maintain the paper as short as possible, the results of other datasets are not shown.}
    \label{fig:recall_results}
\end{figure}

To test the quality of \ac{CF} recommendations of unordered recommendations, precision and recall measures are usually processed, and they are provided separately, or joined in the F1 score. We have done these experiments and we have not found appreciable differences in Recall values for the tested models in the selected datasets. In order to maintain the paper as short as possible, Fig. \Cref{fig:recall_results} only shows the Recall results obtained by processing the \texttt{Movielens 1M} dataset. Results from the rest of datasets are very similar; consequently, the Recall quality measure does not help, in this context, to find out the best \ac{MF} models in the \ac{CF} area.

\begin{figure}[ht]
    \centering
    \includegraphics[width=1.0\textwidth]{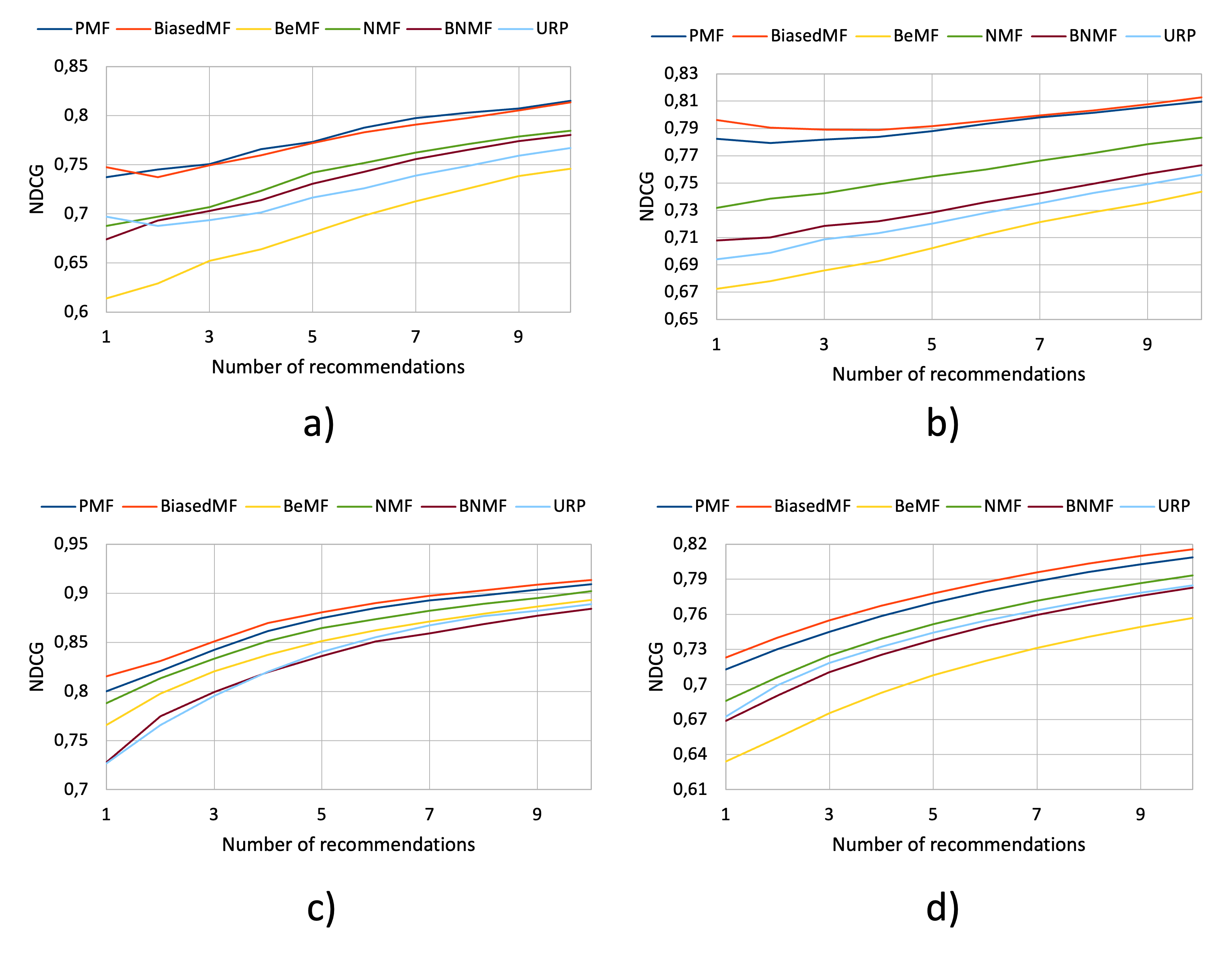}
    \caption{Normalized Discounted Cumulative Gain recommendation quality results; a) \texttt{MovieLens100K}, b) \texttt{MovieLens 1M}, c) \texttt{FilmTrust}, d) \texttt{MyAnimeList}. The higher the values, the better the results.}
    \label{fig:ndcg_results}
\end{figure}

In the \acp{RS} field, recommendations are usually provided in an ordered list. Users' trust in \acp{RS} quickly decays when the first recommendations in the list do not meet their expectations; for that reason, the \ac{NDCG} quality measure particularly penalizes errors in the first recommendations of the list. Fig. \Cref{fig:ndcg_results} (\ac{NDCG} results) shows a similar behavior to Fig. \Cref{fig:precision_results}, where the BiasedMF and \ac{PMF} models provide the best recommendation quality. So, these two models perform fine both in recommending ordered and unordered lists.

\begin{figure}[ht]
    \centering
    \includegraphics[width=1.0\textwidth]{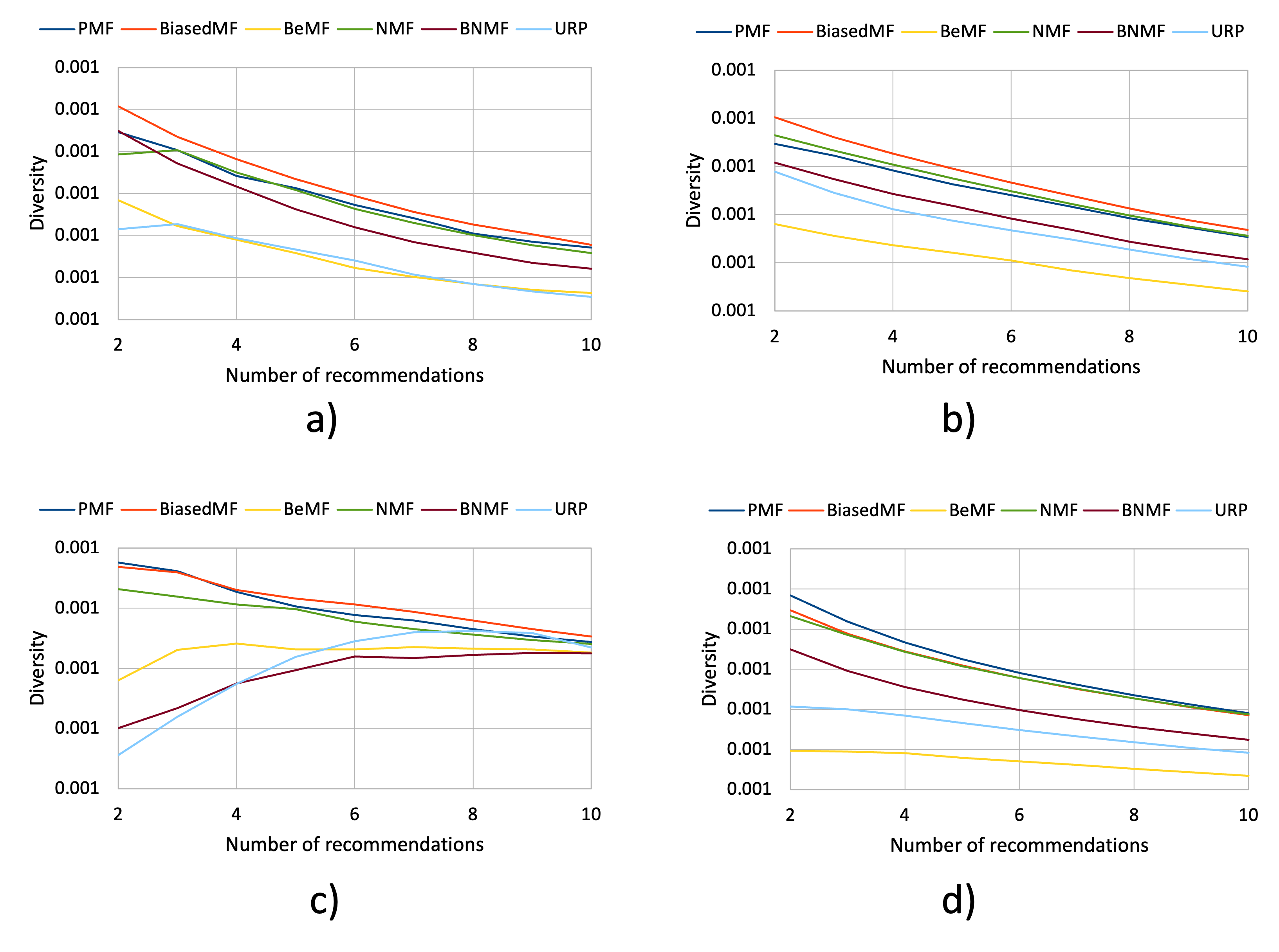}
    \caption{Diversity beyond accuracy results; a) \texttt{MovieLens100K}, b) \texttt{MovieLens 1M}, c) \texttt{FilmTrust}, d) \texttt{MyAnimeList}. The higher the values, the better the results.}
    \label{fig:diversity_results}
\end{figure}

Traditionally, \acp{RS} have been evaluated attending to their prediction and recommendation accuracy; nevertheless, there are some other valuable beyond accuracy aims and their corresponding quality measures. The Diversity measure tests the variety of recommendations, penalizing recommendations focused on the same ‘area’ (Star Wars III, Star Wars I, Star Wars V, Han Solo). Fig. \Cref{fig:diversity_results} shows the Diversity results obtained by testing the selected models; the most diverse recommendations are usually returned when the BiasedMF model is used, followed by both \ac{PMF} and \ac{NMF}. This fact is particularly interesting, since it is not intuitive that the same model (BiasedMF) can, simultaneously, provide accurate and diverse recommendations.

\begin{figure}[ht]
    \centering
    \includegraphics[width=1.0\textwidth]{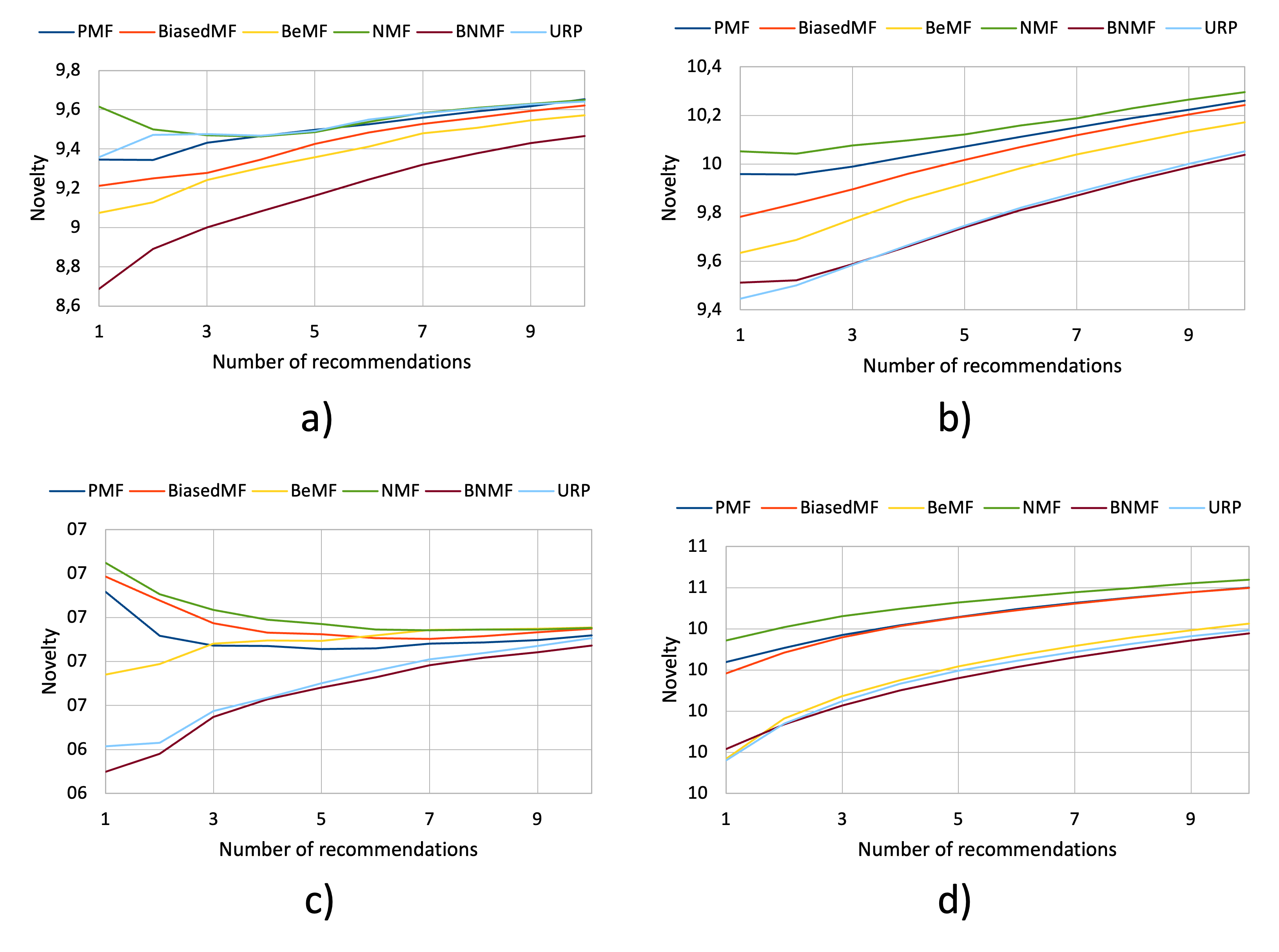}
    \caption{Novelty beyond accuracy quality results; a) \texttt{MovieLens100K}, b) \texttt{MovieLens 1M}, c) \texttt{FilmTrust}, d) \texttt{MyAnimeList}. The higher the values, the better the results.}
    \label{fig:novelty_results}
\end{figure}

Novelty is an important beyond accuracy objective in \acp{RS}. Users appreciate accurate recommendations, but they also want to discover unexpected (and accurate enough) recommendations. Please note that a set of recommendations can be diverse and not novel, as they can be novel and not diverse. It would be great to receive, simultaneously, accurate, novel, and diverse recommendations, but usually improving some of the objectives leads to worsening others. Fig. \Cref{fig:novelty_results} shows the results of the novelty quality measure: \ac{NMF} returns novel recommendations, compared to other models; \ac{NMF} provides a balance between accuracy and novelty.  BiasedMF and \ac{PMF} also provide novel recommendations compared to \ac{BeMF} and \ac{URP}.

\section{Discussion}\label{discussion}

In this section, we provide a comparative discussion of the most adequate \ac{MF} models when applied to a set of different CF databases. To judge each \ac{MF} model, we simultaneously measure a set of conflicting goals: prediction accuracy, recommendation accuracy (unordered and ordered lists) and beyond accuracy aims. We will promote some \ac{MF} models as ‘winners’, attending to their high performance (overall quality results) when applied to the tested datasets. We also provide a summary table to better identify those \ac{MF} models that perform particularly fine on any individual quality objective: novelty, diversity, precision, etc., as well as any combination of those quality measures.

\Cref{tab:comparative} summarizes the results of this section. \ac{BiasedMF} is the most appropriate model when novelty of recommendations is not a particularly relevant issue. \ac{PMF} can be used instead \ac{BiasedMF} when simplicity is required (e.g. educational environments). \ac{BeMF} should only be used when reliability information is required or when reliability values are used to improve accuracy \cite{bobadilla2021neural}. \ac{NMF} and \ac{BNMF} are adequate when semantic interpretation of hidden factors is needed. \ac{NMF} is the best choice when we want to be recommended with novel items. \ac{BNMF} provides good accuracy and it is designed to recommend to group of users. 

\begin{table}[ht]
    \caption{MF models comparative}\label{tab:comparative}%
    \begin{tabular}{@{}lllllll@{}}
        \toprule
         & PMF & BiasedMF & NMF & BeMF & BNMF & URP \\
         \midrule
        MAE & ++ & +++ & + & + & +++ & + \\
        Precision & +++ & +++ & ++ & + & ++ & + \\
        NDCG & +++ & +++ & + & + & + & + \\
        Diversity & ++ & +++ & ++ & + & + & + \\
        Novelty & ++ & ++ & +++ & + & + & + \\
        Total & 12 & 14 & 9 & 5 & 8 & 5 \\
        \botrule
    \end{tabular}%
\end{table}

\section{Conclusions}\label{conclusions}
This paper makes a comparative of relevant \ac{MF} models applied to collaborative filtering recommender systems. Prediction, recommendation, and beyond accuracy quality measures have been tested on four representative datasets. The results show the superiority of the BiasedMF model, followed by the \ac{PMF} one. \ac{BiasedMF} arises as the most convenient model when novelty is not a particularly important feature. \ac{PMF} combines simplicity with accuracy; it can be the best choice for educational or not commercial implementations. \ac{NMF} and \ac{BNMF} are adequate when we want to do a semantic interpretation of their non-negative hidden factors. \ac{NMF} is preferable to \ac{BNMF} when beyond accuracy (novelty and diversity) results are required, whereas it is better to make use of \ac{BNMF} when prediction accuracy is required or when recommending to group of users, or when explaining recommendations is needed. \ac{NMF} and \ac{BiasedMF} are the best choices when beyond accuracy aims are selected, whereas \ac{PMF} or \ac{BiasedMF} performs particularly well in recommendation task, both for unordered and ordered options. \ac{BeMF} can only be selected when reliability values are required or when they are used to improve accuracy.  Finally, \ac{URP} does not seem to be an adequate choice in any of the combinations tested. As future work, it is proposed to add new \ac{MF} models, quality measures, and datasets to the experiments, as well as the possibility of including neural network models such as DeepMF or \ac{NCF}.

\section{Acknowledgments}
This work has been co-funded by the \emph{Ministerio de Ciencia e Innovación} of Spain and European Regional Development Fund (FEDER) under grants PID2019-106493RB-I00 (DL-CEMG) and the \emph{Comunidad de Madrid} under \emph{Convenio Plurianual} with the \emph{Universidad Politécnica de Madrid} in the actuation line of \emph{Programa de Excelencia para el Profesorado Universitario}.

%

\bibliography{biblio}

\end{document}